# Grassmannian beamforming for MIMO-OFDM systems with frequency and spatially correlated channels using Huffman coding

Igor Gutman, Doron Ezri* and Dov Wulich
Ben-Gurion University of the Negev,
Department of Electrical and Computer Engineering
*Communications Laboratory*
Beer-Sheva 84105, ISRAEL,
Tel: ++ 972-8-6472416, Fax: ++ 972-8-6472949,
e-mail: dov@ee.bgu.ac.il

*Greenair Wireless
47 Herut St., Ramat Gan 52551, Israel
Tel: +972-3-6129987
e-mail: doron@greenairwireless.com

*Abstract*

Multiple input multiple output (MIMO) precoding is an efficient scheme that may significantly enhance the communication link. However, this enhancement comes with a cost. Many precoding schemes require channel knowledge at the transmitter that is obtained through feedback from the receiver. Focusing on the natural common fusion of orthogonal frequency division multiplexing (OFDM) and MIMO, we exploit the channel correlation in the frequency and spatial domain to reduce the required feedback rate in a frequency division duplex (FDD) system. The proposed feedback method is based on Huffman coding and is employed here for the single stream case. The method leads to a significant reduction in the required feedback rate, without any loss in performance. The proposed method may be extended to the multi-stream case.

___________________

This work was supported by the Remon Consortium under grant 84680501.

**I. Introduction**

MIMO precoding is an efficient scheme for enhancing the communication link [1], [2]. It may provide better performance than other MIMO techniques without precoding [3]. In most cases the precoding matrix is linked to the singular value decomposition (SVD) [4] of the channel matrix $\mathbf{H}$

$$\mathbf{H} = \mathbf{U}\boldsymbol{\Sigma}\mathbf{V}^{\mathbf{H}} \qquad (1)$$

where $\mathbf{U}$ and $\mathbf{V}$ are unitary matrices and $\boldsymbol{\Sigma}$ is a matrix with real values at diagonal entries, and zero values at other entries. In this paper the matrix $\mathbf{H}$ is a single subcarrier channel matrix. The precoding matrix usually requires information respective to $\mathbf{V}$, or some approximation thereof [1]. While in time division duplex (TDD), the transmitter can estimate the channel matrix $\mathbf{H}$ exploiting channel reciprocity, in FDD feedback from the receiver is required.

In the case of single stream precoding (beamforming), the optimal precoding vector, in the sense of maximal post processing SNR, is the dominant right singular vector of $\mathbf{H}$ (corresponding to the largest singular value) [1]. Thus, in FDD systems that use feedback, the receiver feedbacks a quantized version of the dominant singular vector to the transmitter.

A naive quantization method would use a sufficient number of bits to represent each element of the dominant right singular vector. However, such an approach requires an enormous number of bits to be fed back, which makes it costly to implement in terms of the bit rate of the feedback, so a sophisticated method is required. The *limited feedback* approach [5], [6], [7], is based on quantization of the instantaneous channel state information (CSI) at the receiver. This quantized information is then conveyed to the transmitter using a low-rate feedback channel. Other approaches [8], [9], [10] are based on feedback of statistical channel information, e.g., the channel mean or the channel covariance matrices. These methods, in general, do not perform as well as the ones using instantaneous feedback since they do not track the rapid fluctuations of the channel.

In codebook based methods, the beamforming vector is chosen by the receiver out of a fixed codebook known to both the transmitter and receiver [5], [6], [7]. The codebooks in [5], [6], [7] are designed and optimized for uncorrelated Rayleigh fading channels with various codeword selection criteria based on the instantaneous channel knowledge, so these quantization strategies lack the ability to exploit the existence of the channel correlation. There are codebooks designed for spatially correlated Rayleigh fading channels (using knowledge of the channel spatial correlation matrix) [11], [12], but these works are constrained to a single receiver antenna.

Frequency correlation is usually exploited to reduce the feedback rate by means of grouping and clustering [13]. This approach, however does not fully exploit the frequency correlation, since correlation between clustering is not exploited. None of the above-mentioned techniques provides an algorithm that fully exploits frequency correlation, together with spatial correlation.

In this paper, we propose a new method that exploits frequency and spatial correlation to reduce the feedback rate. The method is based on a differential codebook approach and Huffman coding. The codeword length in our method is variable and is a function of the difference between the indexes[1] of precoding vectors at neighboring clusters. Although the proposed method is designed for frequency correlated channels, it also provides feedback rate reduction for spatially-correlated channels. Our method may be viewed as lossless compression of the required feedback in systems employing a Grassmannian codebook. Thus, no performance loss

---

[1] Later we will define an index that represents each one of the code words that assembles the codebook.

is introduced and the precoding gains are exactly as with classical Grassmannian beamforming [6].

The rest of this paper is organized as follows. Section 2 lays out the model of the system and the problem statement. Section 3 presents a differential setup derived from the Grassmannian codebook. Section 4 presents the proposed algorithm based on the differential setup and Huffman coding. Simulations results are given in Section 5 and finally, conclusion is discussed in Section 6.

**II. Overview**

*II.1 Grassmannian beamforming*

A baseband equivalent system of a single MIMO-OFDM subcarrier is shown in Fig. 1. There are $M_t$ and $M_r$ transmit and receive antennas, respectively, and the total number of subcarriers at OFDM symbol is $L$. It is assumed that the channel matrix $\mathbf{H}(l)$, $l = 1, 2, ..., L$ at the $l$-th subcarrier has Gaussian entries, i,e, $h_{rs}(l) \sim CN(0,1)$, $r = 1, 2, ..., M_r$, $s = 1, 2, ..., M_t$. It is also assumed that frequency correlation between subcarriers exists, which is described by the frequency correlation matrix

$$\mathbf{R}_f = E\{\mathbf{h}_{rs} \cdot \mathbf{h}_{rs}^H\}, \qquad (2)$$

where $\mathbf{h}_{rs} \stackrel{def}{=} [h_{rs}(1), h_{rs}(2), ..., h_{rs}(L)]^T$. We assume here that $\mathbf{R}_f$ is independent of $s$ and $r$.

In single stream beamforming, the steering vector at the transmitter is the dominant right singular vector $\mathbf{v}$ of the single subcarrier channel matrix $\mathbf{H}$ [1], [6]. The matrix $\mathbf{H}$ is estimated at the receiver after which the vector $\mathbf{v}$ is calculated using SVD. In cases, when the FDD mode is being implemented, the vector $\mathbf{v}$ has to be fed back to the transmitter using a feedback channel. It is assumed that this feedback channel is instantaneous and error free but has a limited speed. To meet the condition of limited speed, the vector $\mathbf{v}$ must be quantized using a codebook $\mathbf{W}$. Here we use a codebook of size $N$ according to the Grassmannian line packing (GLP). The codewords $\mathbf{v}_i \in \mathbf{W}, i=1,2,...,N$ are chosen to maximize the minimum distance between codewords [6]

$$J = \min_{1 \leq k,l \leq N} \sqrt{1 - \left|\mathbf{v}_k^H \mathbf{v}_l\right|^2}, \quad (3)$$

which results in a natural GLP solution, thanks to the isotropicity[2] of the maximal right singular vector.

*II.2 The OFDM-MIMO channel and frequency clustering*

We assume that: (i) the delay spread of the channel results in frequency correlation between the OFDM subcarriers and (ii) the close spacing between the TX and/or RX antennas (among other factors) results in spatial correlation. For the case of a frequency correlation between the subcarriers, Mondahl and Heath [13] proposed to group a number of neighboring subcarriers into clusters, where a single index, representing a single codeword, is fed back for each cluster. Out of the several criteria of cluster precoding presented by Mondahl and Heath, we adopt the one that maximizes the minimum SNR within the cluster. Let $\mathbf{v}_i(m)$ be the optimal codeword with index $i$ of cluster $m$ chosen according to:

---

[2] According to [6] the line generated by the optimal beamforming vector for MIMO Rayleigh fading channel is an isotropically oriented line in $\mathbf{C}^{M_t}$ passing through the origin. Therefore, the problem of quantized transmit beamforming in a MIMO communication system reduced to quantizing an isotropically oriented line in $\mathbf{C}^{M_t}$.

$$\mathbf{v}_i(m) = \arg\left\{\max_{\substack{i=1...N \\ \mathbf{v}_i \in \mathbf{W}}} \left[\min_{1 \leq g \leq G} \|\mathbf{H}[(m-1)\cdot G + g]\cdot \mathbf{v}_i\|_2^2\right]\right\}, \ m=1,2,...,M = \frac{L}{G}, \quad (4)$$

where $G$ denotes the number of subcarriers in the cluster. Note however that all clustering methods proposed by Mondahl and Heath do not fully exploit the potential of frequency correlation because they assume that there is no correlation between the clusters.

One of the purposes of this paper is to propose a method that fully exploits the frequency correlation between the clusters. The proposed method does not depend on any specific codebook nor on a specific clustering criterion. Rather it depends on the ability to use the frequency and spatial correlations.

### III. Differential setup

In case of frequency correlation between clusters, it is expected that the codebook indexes of nearby clusters may also exhibit correlation. In order to exploit this correlation we define a differential setup for the codebook indexes. Specifically, we address the probability of each index and transition probabilities

*III.1. Transition probabilities*

For each of the codewords $\mathbf{v}_i \in \mathbf{W}, \ i=1,2,...,N$ we define

$$\alpha_i(l) \stackrel{def}{=} \left|\mathbf{v}_i^H \mathbf{v}_l\right|^2, \ l=1,2,...,N \quad (5)$$

and arrange the $\alpha_i(l)$ in descending order resulting in $\beta_i(k)$. Thus we have:

(i) $$\beta_i(1) = \alpha_i(i) = 1,$$

and

(ii) $$\beta_i(k+1) \leq \beta_i(k), \ k=1,2,...,N. \quad (6)$$

If for some $k$ there is an equality in (6) we arrange these $\beta_i$ in an arbitrary order. As a result, for each of the vectors $\mathbf{v}_i$ there exists a set of vectors $\mathbf{q}_i(k)$ such that:

$$\left|\mathbf{v}_i^H \mathbf{q}_i(k)\right|^2 = \beta_i(k), \qquad (7)$$

Table 1 illustrates the result of the ordering described above for a particular Grassmannian codebook of size $N = 64$ for $M_t = 4$, as given in [14].

$i \longrightarrow$

Table 1

$k \downarrow$

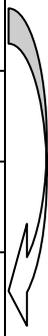

| $\mathbf{v}_1$ | | $\mathbf{v}_2$ | | ... | $\mathbf{v}_{64}$ | |
|---|---|---|---|---|---|---|
| $\mathbf{q}_1(k)$ | $\beta_1(k)$ | $\mathbf{q}_2(k)$ | $\beta_2(k)$ | | $\mathbf{q}_{64}(k)$ | $\beta_{64}(k)$ |
| $\mathbf{q}_1(1)=\mathbf{v}_1$ | $\beta_1(1)=1$ | $\mathbf{q}_2(1)=\mathbf{v}_2$ | $\beta_2(1)=1$ | | $\mathbf{q}_{64}(1)=\mathbf{v}_{64}$ | $\beta_{64}(1)=1$ |
| $\mathbf{q}_1(2)=\mathbf{v}_4$ | $\beta_1(2)=0.81$ | $\mathbf{q}_2(2)=\mathbf{v}_{63}$ | $\beta_2(2)=0.81$ | | $\mathbf{q}_{64}(2)=\mathbf{v}_{61}$ | $\beta_{64}(2)=0.81$ |
| $\mathbf{q}_1(3)=\mathbf{v}_{62}$ | $\beta_1(3)=0.81$ | $\mathbf{q}_2(3)=\mathbf{v}_5$ | $\beta_2(3)=0.81$ | | $\mathbf{q}_{64}(3)=\mathbf{v}_3$ | $\beta_{64}(3)=0.81$ |
| $\mathbf{q}_1(4)=\mathbf{v}_{20}$ | $\beta_1(4)=0.78$ | $\mathbf{q}_2(4)=\mathbf{v}_{21}$ | $\beta_2(4)=0.78$ | | $\mathbf{q}_{64}(4)=\mathbf{v}_{45}$ | $\beta_{64}(4)=0.78$ |
| $\mathbf{q}_1(5)=\mathbf{v}_{46}$ | $\beta_1(5)=0.78$ | $\mathbf{q}_2(5)=\mathbf{v}_{47}$ | $\beta_2(5)=0.78$ | | $\mathbf{q}_{64}(5)=\mathbf{v}_{19}$ | $\beta_{64}(5)=0.78$ |
| . | . | . | . | | . | . |

Property 1. For any $i$ and $j$

$$\beta_i(k) = \beta_j(k), \ k = 1, 2, ..., N. \qquad (8)$$

Equation (8) follows directly from the properties of the GLP [6].

Suppose that the codeword of some cluster, say cluster $m$, is $\mathbf{v}_i$ while the codeword of the neighbor cluster, $m+1$, is $\mathbf{v}_j$, $j=1,2,...,N$. In this case we define a random variable

$$X(m+1)_i \stackrel{def}{=} \xi_i(j), \qquad (9)$$

where $\xi_i(j)$ is a function that describes the amounts of indexes according table 1 between $i$ and $j$. To illustrate the meaning of the r.v. $X(m)_i$ let us consider the following example based on Table 1.

Example Suppose that the codeword index of cluster $m$ is $i=2$ and the index of the codeword of the neighboring cluster $m+1$ is $j=21$. In Table 1 in column of $\mathbf{v}_2$, column two, we have $\xi_2(21)=4$; therefore $X(m+1)_2=4$.

□

Property 2. The distribution of $X(m)_i$ does not depend on $i$ for spatially uncorrelated channels due to the isotropicity of the maximal right singular vector, i.e.

$$\Pr(X(m)_i = k) = p(k)_i = p(k). \qquad (10)$$

Equation (10) follows directly from the properties of the GLP. As a result of property 2 we may disregard the index $i$ in $X(m)_i$ for spatially uncorrelated channels. For spatially correlated channels however, the distribution of the maximum right singular vector is not isotropic, and therefore (10) is not valid $\left(p(k)_i \neq p(k)_j\right)_{i \neq j}$.

Now consider $M$ clusters spanned in the frequency axis and their respective codewords: $\{\mathbf{v}(m)_i\}_{m=1}^{M}$. Let $X(m)_i$ denote the random variable defined according to (9), $X(m+1)_i$ indicates how close $\mathbf{v}(m+1)_i$ is to $\mathbf{v}(m)_j$.

For frequency-correlated clusters, the probabilities $p(k)_i$ are not equal; therefore it is expected that the Huffman coding of the set of r.v. $\{X(m)_i\}_{m=2}^{M}$ will result in a significant reduction of the feedback bit rate. Consider, for example, the case of a low frequency correlation (and no spatially uncorrelation), as shown in Fig. 3, and the Grassmannian codebook taken from [14] for $N=64$ and $M_t=4$. For $G=1$, $p(1)=0.35$ while $p(k) \leq 0.05$, $k=2,3,...,N$. Therefore a significant reduction in the feedback bit rate is expected for such a probability distribution.

*III.2. The impact of spatial correlation on transition probabilities*

Besides the fact that for spatially correlated channels the transition probabilities $p(k)_i$ may depend on $i$, the spatial correlation also affects the transition probabilities distribution for $k$. The transmitter performs beamforming using the dominant right singular vector, i.e., the singular vector with the largest eigen-value. However, since channel matrices $\mathbf{H}(l)$ are chosen at random, an eigen-value switching[3] may occur. The probability of such an event depends on the spatial correlation (among other factors like frequency correlation) and is high when there is no spatial correlation (low condition number), and decreases when there is high spatial correlation[4] (high condition number) see example at Fig. 4a and Fig. 4b. Eigen-mode switching causes an increase of $\{p(k)_i\}_{k,i=1}^{N}$ for large $k$ which in turn increases the entropy of the r.v. $X$. This results in an increase number of Huffman coding bits.

---

[3] Eigen-value switching occurs when there is a different eigen mode at neighbor frequency clusters, at this case, even though high frequency correlation between clusters, there is no correlation at all between dominant right singular vectors of neighbor clusters.

[4] The spatial correlation depends (among other factors) on the ratio between the distance of the antennas and the wavelength. If the value of this ratio decreases then the probability of eigen-mode switching increases.

## IV The Huffman coding based algorithm

*IV.1. The algorithm*

We consider a scenario where subcarriers are clustered into $M$ clusters. Assuming that the first cluster is fully encoded using $\log_2 N$ bits, the consecutive clusters may be encoded by encoding the random variables $\{X(m)_i\}_{m=2}^{M}$ instead of the codeword $\mathbf{v}(m)_i$. As a result, the block message

$$[X(1)_i, X(2)_j, ..., X(M)_k] \qquad (11)$$

per OFDM symbol must be fed back to the transmitter. Recall also that $X(1)_i$ denotes the index $i$ of the codeword of $\mathbf{v}(1)_i$. $X(1)_i$ is transmitted as is, while $[X(2)_j, ..., X(M)_k]$ are encoded using the Huffman algorithm.

If the probabilities $\{p_k\}_{k=1}^{N}$ are not correct, the average number of bits due to Huffman coding is greater than the entropy. Moreover, if the incorrect probabilities are known to both the transmitter and the receiver, then the side which receives the information via the feedback channel (the transmitter in our case), will decode the message without errors thereby causing no degradation in performance. In this case the rate of the feedback is greater than the entropy of $[X(1)_i, X(2)_j, ..., X(M)_k]$.

The fact that the receiver and the transmitter share the same information even if the probabilities are incorrect provides a way to estimate $\{p(k)_i\}_{k,i=1}^{N}$ "on the fly" and simultaneously by both the transmitter and the receiver. It should be noted that for error-free feedback, the estimates done by the transmitter and the receiver are exactly the same. Figure 2a shows the flow chart of the receiver's (the side which transmitts the feedback side) state machine and Fig. 2b shows the flow chart of the transmitter's (the side which receives the feedback) state machine.

*IV.2. Updating the transition probabilities*

Assuming that the frequency correlation properties of the channel do not change in time, the sample size of the statistic to estimate $\{p(k)_i\}_{k,i=1}^{N}$ increases as the number of OFDM symbols increases. Each OFDM symbol supplies $M-1$ realizations of the r.v. $X$.

For spatial uncorrelated channel, once $T$ OFDM symbols have been transmitted, the probabilities $\{p(k)\}_{k=1}^{N}$ are approximated by:

$$\hat{p}(k) = \frac{number\ of\ realizations = k}{(M-1)\cdot T}. \tag{12}$$

As is widely known, $p_k = \lim_{M\cdot T \to \infty} \hat{p}_k$, the result of which is that the number of bits of Huffman coding will approach the entropy of $X$.

In case of spatially correlated channel, in order to estimate the probabilities $\hat{p}(k)_i$, the index *i* of codeword should be taken into consideration.

The main drawback for assuming a spatial correlated channel is slower convergence time of transition probabilities to the true ones, since the number of realizations to estimate $\hat{p}(k)_i$ for each index *i* is less then $M-1$ for each OFDM symbol.

From simulations that we performed, it is clearly seen that for spatial correlated channels $\left(p(k)_i \approx p(k)_j\right)_{i \neq j}$, and very close to (10), therefore in section 5, we assume for the algorithm that the channel is spatially uncorrelated(even when the actual channel is a spatial correlated), and therefore avoiding slow convergence process. This cause the feedback cost gain result to be sub-optimal (only when the channel is spatially correlated), however still very impressive. In case of spatially uncorrelated channel, the algorithm is optimal.

## V. Performance analysis

### V.1. Statistical channel modeling

To analyze the performance of the proposed algorithm we chose an OFDM with $L=64$ subcarriers, $M_t=4$ and $M_r=2$. A 6-bit Grassmannian codebook [14] was used. We consider a case of low and high frequency correlations. Figure 3 shows the correlation function of the low (solid line) and high (dashed line) correlations. In addition to the frequency correlation we consider both zero and high spatial correlation. For the latter we consider the TX spatial correlation matrix

$$\mathbf{R}_t = \begin{pmatrix} 1 & a & b & c \\ a^* & 1 & a & b \\ b^* & a^* & 1 & a \\ c^* & b^* & a^* & 1 \end{pmatrix} \qquad (13a)$$

and the RX spatial correlation matrix

$$\mathbf{R}_r = \begin{pmatrix} 1 & a \\ a^* & 1 \end{pmatrix} \qquad (13b)$$

with $a = 0.4640 + 0.8499j$, $b = -0.4802 + 0.7421j$, $c = -0.7688 - 0.0625i$. This scenario, taken from 3GPP, corresponds to a macro-cell scenario with a Laplacian power azimuth spectrum and an azimuth spread of 5 degrees [15].

To model the frequency and spatial correlation we introduce 3-D channel matrix $\mathbf{H}_L$ with dimension $N_r \times N_t \times L$. Its correlation matrix which includes the frequency and spatial correlations is given by [16]

$$\mathbf{R} = \mathbf{R}_f \otimes (\mathbf{R}_t \otimes \mathbf{R}_r), \qquad (14)$$

where $\otimes$ denotes the Kronecker product and $\mathbf{R}_f$ is the frequency correlation matrix according to Fig. 3. Using the eigenvalue decomposition of $\mathbf{R} = \mathbf{P}\Omega\mathbf{P}^H$ the random 3D channel matrix is generated according to:

$$\mathbf{H}_L = devec\left(\mathbf{P}\sqrt{\Omega}\cdot\mathbf{x}\right), \qquad (15)$$

where the vector $\mathbf{x}$ is an i.i.d. (Gaussian) vector.

*V.2. Performance*

The performance of the proposed algorithm depends on the clustering of the subcarriers. Table 2 and Table 3 show the total number of bits per OFDM symbol as a function of $G$ (the number of subcarriers per cluster) that must be fed back to the transmitter. Both kinds of spatial correlation are considered. It is assumed that the probabilities $\{p(k)\}_{k=1}^{N}$ are exactly known.

Table 2. Low frequency correlation

| $G$ | 1 | 2 | 4 |
|---|---|---|---|
| Number of bits for clustering only | 384 | 192 | 96 |
| Number of bits for Huffman coding without spatial correlation | 294.1 | 184.1 | 95.7 |
| Number of bits for Huffman coding with spatial correlation | 146 | 85.2 | 45.5 |

Table 3. High frequency correlation

| $G$ | 1 | 2 | 4 | 8 | 16 |
|---|---|---|---|---|---|
| Number of bits for clustering only | 384 | 192 | 96 | 48 | 24 |
| Number of bits for Huffman coding without | 113 | 80.4 | 61.4 | 40.1 | 23 |
| spatial correlation | | | | | |
| Number of bits for Huffman coding with spatial correlation | 83.2 | 50.9 | 32.9 | 21.8 | 13.6 |

Figures 5a and 5b respectively show the number of bits of Huffman coding as a function of $T$ for the case of a low frequency correlation without spatial correlation and for a high frequency correlation with spatial correlation. It is clearly visible that the system is in steady state after 10 OFDM symbols for a low frequency correlation have been transmitted and after 30 OFDM symbols for a high frequency correlation have been transmitted.

## VI. Conclusion

A method based on Grassmannian beamforming and Huffman coding to reduce the feedback speed in frequency and spatially correlated channels in MIMO-OFDM was proposed. The method fully exploits the (frequency) correlation between the neighboring subcarriers/clusters. Although it was designed for frequency correlated channels, it also reduces the feedback rate for spatially correlated channels, however the main drawback when we assume spatially correlated channels, is slower convergence time of transition probabilities. The problem of slow convergence time can be solved by using a sub-optimum solution by assuming that the actual channel is spatially uncorrelated.

The examples show that the highest feedback reduction rate is obtained for the case of no clustering. The feedback rate is reduced by a factor 1.3 for low frequency correlation and by a factor of 3.4 for a high frequency correlation for channels with no spatial correlation. For high spatial correlation, the feedback rate reduction is increased to 2.6 and 4.6 for low and high frequency correlation, respectively.

The transition probabilities update takes at most 30 OFDM symbols for OFDM with 64 subcarriers which make the proposed method very suitable for most practical cases when the correlations between subcarriers remain constant for many OFDM symbols.

We considered at this work the case of single stream beamforming. However the same principle of coding may be extended to multi-stream precoding, where precoding matrices are employed, and similar gains in feedback rate are expected.

**References**

[1] A. J. Paulraj, D. A. Gore, R. U. Nabar, and H. Bolcskei, "An Overview of MIMO Communications – A Key to Gigabit Wireless", *Proceeding of IEEE*, Vol. 92, issue 2, pp. 198-218, Feb 2004.

[2] I. E. Telatar, "Capacity of Multi-antenna Gaussian Channels", *European Transactions on Telecommunications*, Vol. 10, No. 6, pp. 585-595, Nov, 1999.

[3] A. Hottinen, O. Tirkkonen and T. Wichman, *Multi-antenna Transceiver Techniques for 3G and Beyond*, Wiley, 2003, pp. 209-228.

[4] G. Lebrun, J. Gao and M. Faulkner, "MIMO Transmission Over a Time-Varying Channel Using SVD", *IEEE Transactions on Wireless Communications,* Vol. 4, No. 2, pp. 757-764, March 2005.

[5] K. K. Mukkavilli, A. Sabharwal, E. Erkip and B. Aazhang, "On Beamforming with Finite Rate Feedback in Multiple-Antenna Systems", *IEEE Transactions on Information Theory*, Vol. 49, No. 10, October, 2003.

[6] D. J. Love, R. W. Heath, Jr., and T. Strohmer, "Grassmannian Beamforming for Multiple-Input Multiple Output Wireless Systems", *IEEE Transactions on Information Theory*. 49, pp. 2735 – 2747, Oct. 2003.

[7] A. Narula, M. J. Lopez, M. D. Trott, and G. W. Wornell, "Efficient Use of Side Information in Multiple-Antenna Data Transmission over Fading Channels," IEEE Jour. Select. Areas in Commun., vol. 16, no. 8, pp. 1423–1436, Oct. 1998.

[8] B. C. Banister and J. R. Zeidler, "A Simple Gradient Sign Algorithm for Transmit Antenna Weight Adaptation with Feedback," IEEE Trans. Sig. Proc., vol. 51, no. 5, pp. 1156–1171, May 2003.

[9] J. C. Roh and B. D. Rao, "Transmit Beamforming in Multiple-Antenna Systems with Finite Rate Feedback: A VQ-Based Approach," IEEE Trans. Inform. Theory, vol. 52, no. 3, pp. 1101–1112, Mar. 2006.

[10] B. Mondal and R. W. Heath, Jr., "Channel Adaptive Quantization for Limited Feedback MIMO Beamforming Systems," IEEE Trans. Signal Processing, vol. 54, no. 12, pp. 4731–4740, Dec. 2006.

[11] P. Xia and G. B. Giannakis, "Design and Analysis of Transmit Beamforming based on Limited-Rate Feedback," IEEE Trans. Sig. Proc., vol. 54, no. 5, pp. 1853–1863, May 2006.

[12] D. J. Love and R.W. Heath, Jr., "Limited Feedback Diversity Techniques for Correlated Channels," IEEE Trans. Veh. Tech., vol. 55, no. 2, pp. 718–722, Mar. 2006.

[13] B. Mondahl and R. W. Heath Jr ., "Algorithms for Quantized Precoding in MIMO OFDM Beamforming Systems", *Proceedings of the SPIE*, Volume 5847, pp. 80-87 (2005).

[14] "Proposed Text of MIMO for the IEEE 802.16m Amendment" *IEEE 802.16 Broadband Wireless Access Working Group*, http://ieee802.org/16, March 2009.

[15] L. Schumacher, "Description of the MATLAB(R) Implementation of a MIMO Channel Model Suited for Link-Level Simulations," Available from http://www.istimetra.org/ , March 2002, version 0.1.

[16] W. Weichselberger, H. O¨ zcelik, M. Herdin, and E. Bonek, "A Novel Stochastic MIMO Channel Model and its Physical Interpretation," in *Proc. International Symposium on Wireless Personal Multimedia Communications, WPMC,* Yokosuka, Japan, October 2003.

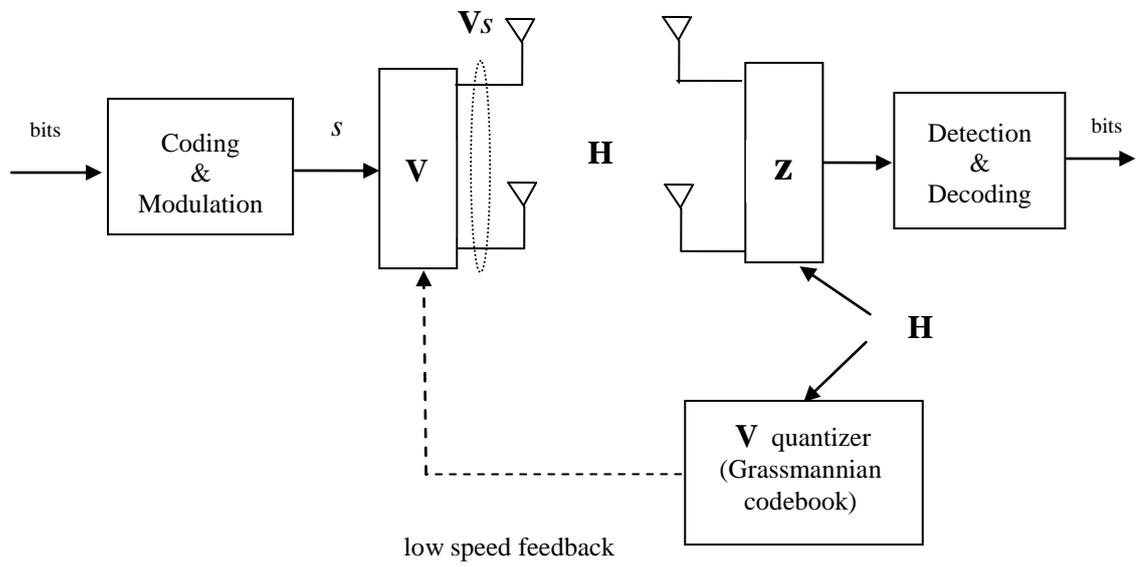

Fig. 1. Baseband equivalent system of single MIMO-OFDM subcarrier

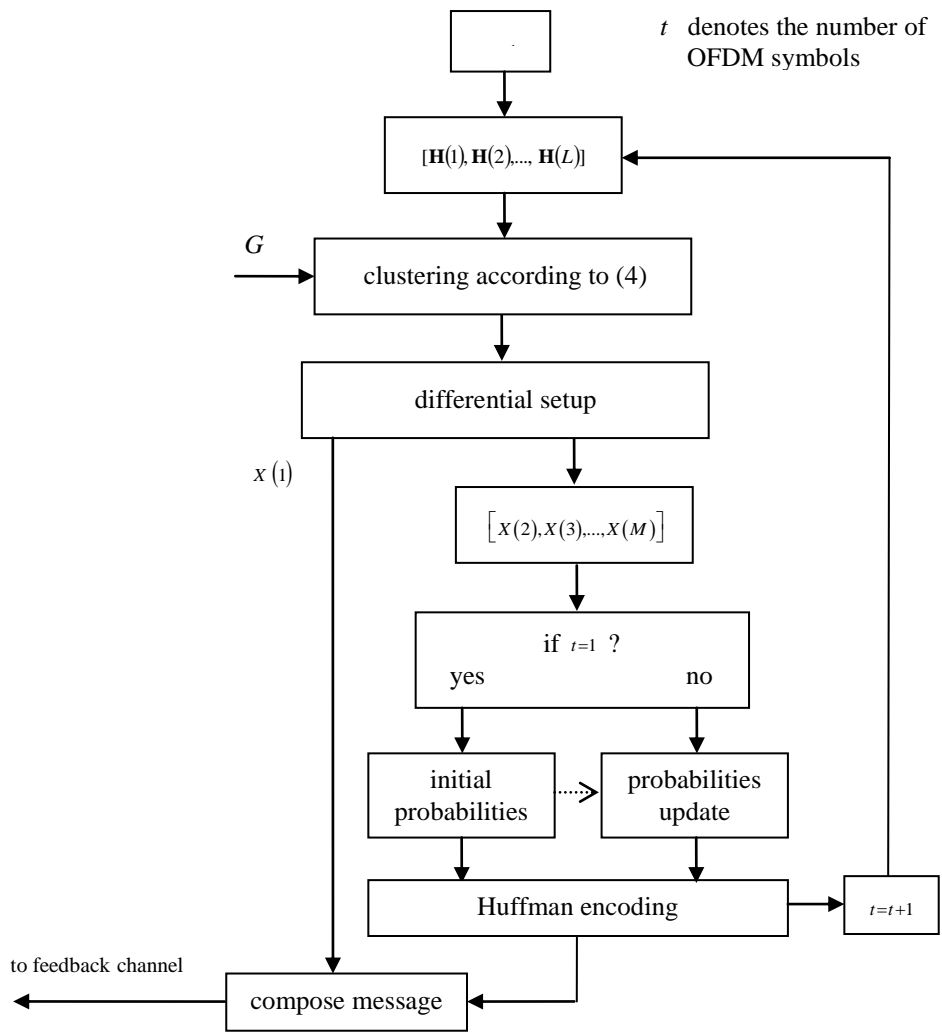

Fig. 2a. Flow chart of the state machine of the receiver

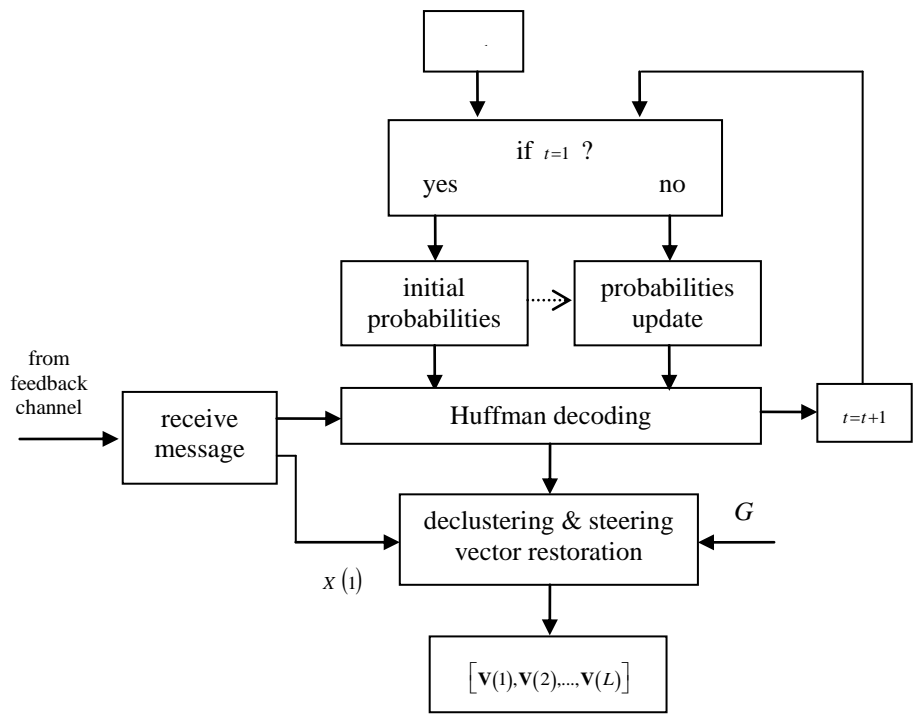

Fig. 2b. Flow chart of the state machine of the transmitter

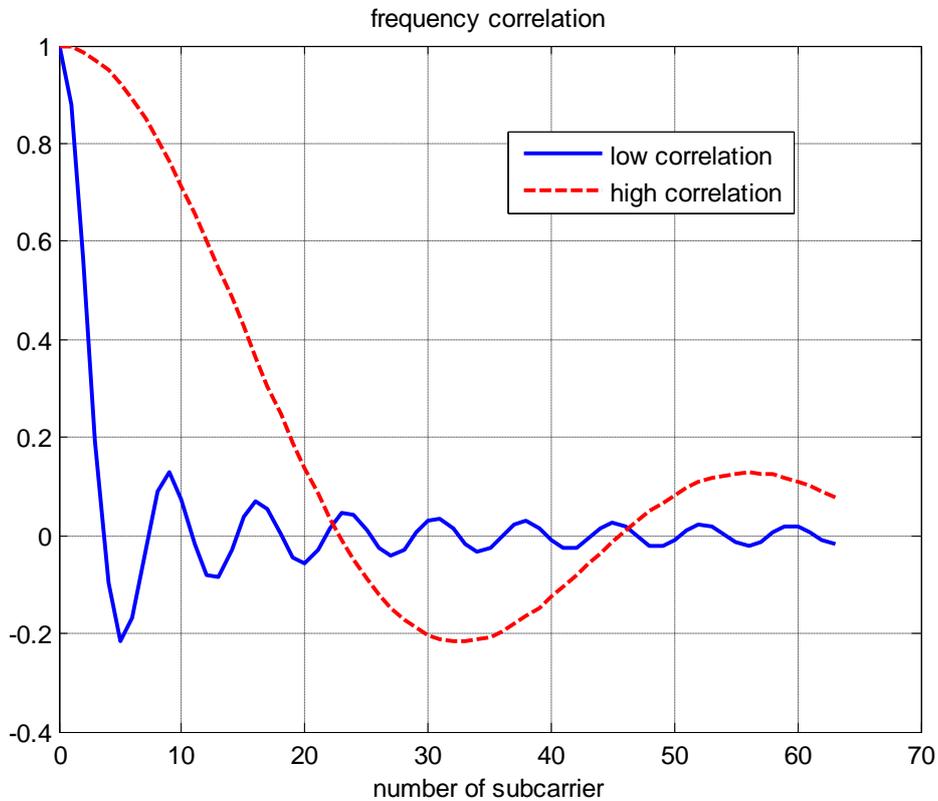

Fig. 3. Illustration of low and high frequency correlation

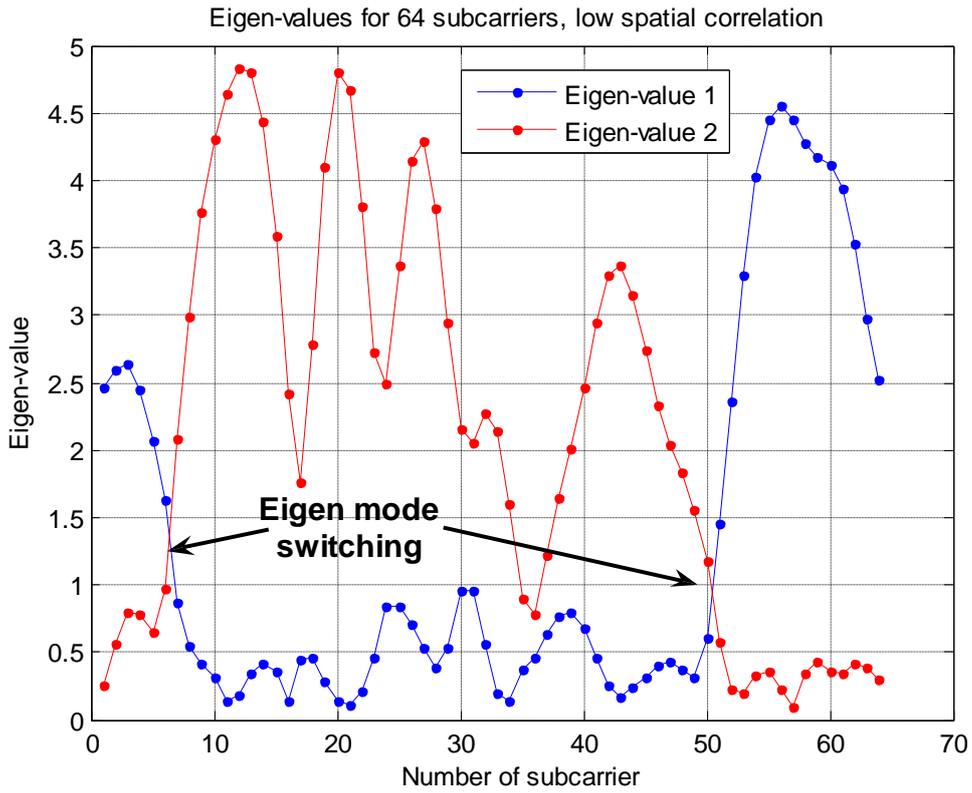

Fig. 4a. Eigen-values for the case of no spatial correlation

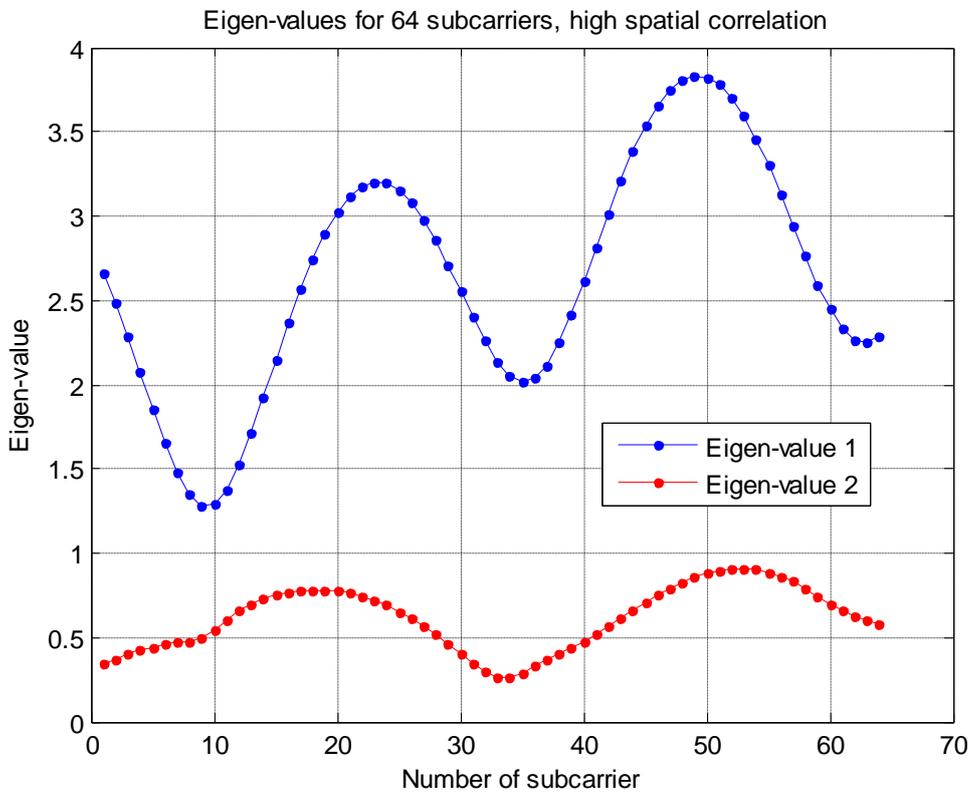

Fig. 4b. Eigen-values for the case of high spatial correlation

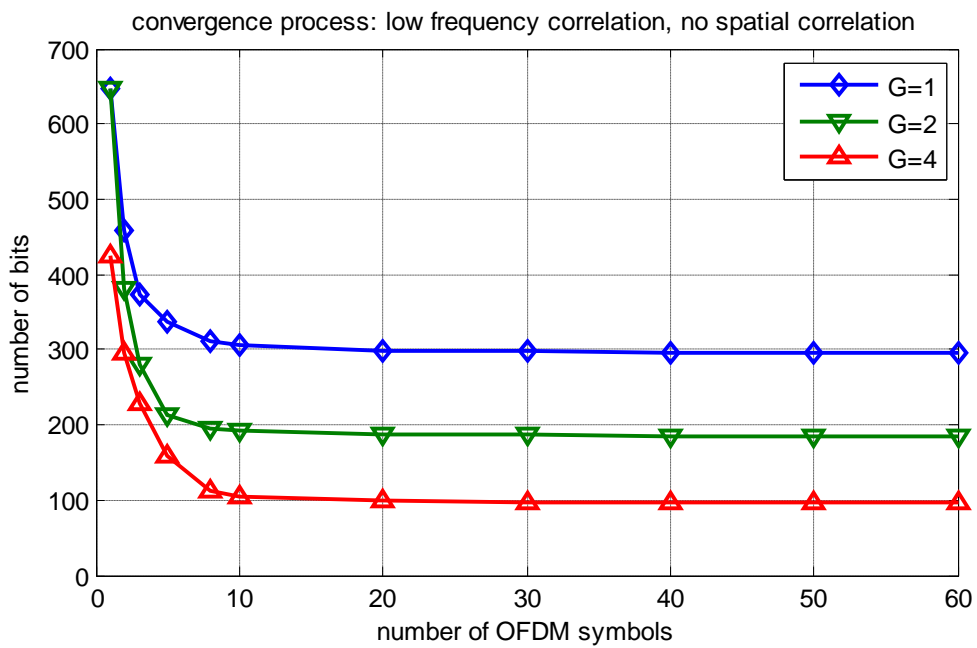

Fig. 5a. Number of Huffman coding bits vs. number of OFDM symbols

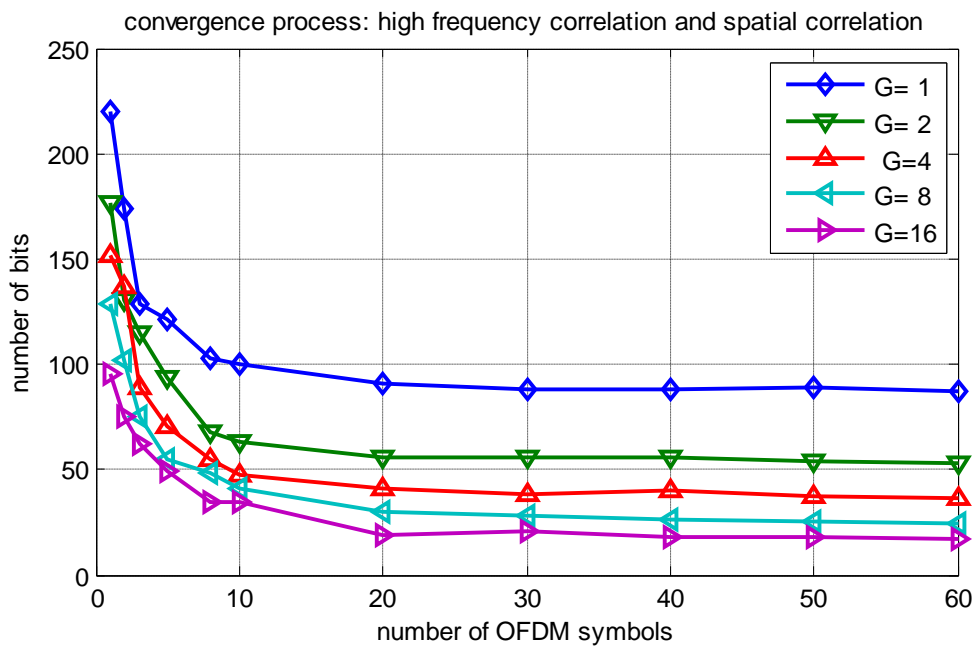

Fig. 5b. Number of Huffman coding bits vs. number of OFDM symbols